\documentclass[aps,prl, reprint,amsmath,amssymb, superscriptaddress]{revtex4-1}
\usepackage{graphicx} 
\usepackage{amssymb,amsfonts,amsmath}
\usepackage{txfonts}
\usepackage{xcolor}              
\usepackage{setspace}            
\usepackage{fullpage}            
\usepackage{url}
\usepackage[caption=false]{subfig}
\usepackage{booktabs}
\usepackage{url}
\usepackage{placeins}
\usepackage{epstopdf}
\usepackage{ragged2e}
\usepackage{lipsum}

\definecolor{darkgreen}{HTML}{00BB00}

\def\er{Erd\H{o}s-R\'enyi }

\begin{document}
\title{Critical field-exponents  for secure message-passing in modular networks}
\date{\today}
\author{Louis M. Shekhtman}
\affiliation{ Department of Physics, Bar-Ilan University, Ramat Gan, Israel}
\author{Michael M. Danziger}
\affiliation{ Department of Physics, Bar-Ilan University, Ramat Gan, Israel}
\author{Ivan Bonamassa}
\affiliation{ Department of Physics, Bar-Ilan University, Ramat Gan, Israel}
\author{Sergey Buldyrev}
\affiliation{ Department of Physics, Yeshiva University, New York, USA}
\author{Guido Caldarelli}
\affiliation{IMT Alti Studi Lucca, Piazza San Francesco 19, 55100 Lucca Italy}
\affiliation{CNR-ISC Dipartimento di Fisica, University of Rome Sapienza, Piazzale Aldo Moro 2, 00185 Rome Italy}
\affiliation{London Institute for Mathematical Sciences, 35a South Street, Mayfair London UK}
\affiliation{European Centre for Living Technology (ECLT)  Ca' Foscari San Marco 2940-30124 Venezia , Italy}
\author{Vinko Zlati\'{c}}
\affiliation{CNR-ISC Dipartimento di Fisica, University of Rome Sapienza, Piazzale Aldo Moro 2, 00185 Rome Italy}
\affiliation{ Theoretical Physics Division, Institute ``Ruder Boskovic", Zagreb, Croatia}
\author{Shlomo Havlin}
\affiliation{ Department of Physics, Bar-Ilan University, Ramat Gan, Israel}

\begin{abstract} 
   We study secure message-passing in the presence of multiple adversaries in modular networks. We assume a dominant fraction of nodes in each module have the same vulnerability, i.e., the same entity spying on them. We find both analytically and via simulations that the links between the modules (interlinks) have effects analogous to a magnetic field in a spin system in that for any amount of interlinks the system no longer undergoes a phase transition. We then define the exponents $\delta$, which relates the order parameter (the size of the giant secure component) at the critical point to the field strength (average number of interlinks per node), and $\gamma$, which describes the susceptibility near criticality. These are found to be $\delta=2$ and $\gamma=1$ (with the scaling of the order parameter near the critical point given by $\beta=1$). When two or more vulnerabilities are equally present in a module we find $\delta=1$ and $\gamma=0$ (with $\beta\geq2$). Apart from defining a previously unidentified universality class, these exponents show that increasing connections between modules is more beneficial for security than increasing connections within modules. We also measure the correlation critical exponent $\nu$, and the upper critical dimension $d_c$, finding that $\nu d_c=3$ as for ordinary percolation, suggesting that for secure message-passing $d_c =6$. These results provide an interesting analogy between secure message-passing in modular networks and the physics of magnetic spin-systems.
\end{abstract}
\maketitle
As our world becomes more interconnected, the need to pass messages securely has gained increasing importance \cite{katz2014introduction}. The recently developed applications of statistical physics of networks to anonymous browsing networks \cite{de2017modeling}  and secure message-passing \cite{krause2016hidden} promises an  interesting new direction of security based on network topology.  
One application is internet routers, which form a physical communication network with nodes belonging to specific countries that can eavesdrop on information passing through their routers \cite{beimel2011secret}.   
Whether information can be transferred through such a communication network securely and effectively is strongly dependent on the frequency and structural network properties of vulnerabities e.g. nodes belonging to a malicious country in the aforementioned example. In this Letter we generalize the framework of ``color-avoiding percolation" (CAP) \cite{krause2016hidden,krause2016color} to study a more realistic case of secured message passing in a communication network with a given \emph{community structure} and different classes of adversaries (vulnerabilities). 

In CAP each node in the network is assigned a specific color. A path between two nodes is considered to avoid a particular color (i.e., is secure from that color) if no nodes of that color exist along the path (not counting its endpoints). If between two given nodes there is for each color at least one path avoiding that color, the two nodes are considered securely connected. Equivalently, only nodes that can communicate such that no single color exists on every path between them are considered secure. 

Here we consider CAP on networks with given community structure, a realistic case for many networks \cite{newman2004finding,*girvan2002community,mucha2010community,*porter2009communities,radicchi2004defining,shai2015critical, shekhtman2015resilience,porter2009communities,dorogovtsev2008organization,lancichinetti2009detecting}. Continuing the above example of internet routers, in each country most of the routers presumably belong to that country with a smaller number of routers belonging to other countries \cite{eriksen2003modularity,krause2016hidden,vazquez2002large}. To study the community structure we use the stochastic block model \cite{fienberg1985statistical, peixoto2013parsimonious}, where each community is recognized as a `block' in an adjacency matrix,  and assign a certain color to dominate each module. This imposes correlations on the distribution of colors in the network, naturally modeled as a modular network. %

\begin{figure}[h]
	\centering
	\vfill
	\subfloat{%
		\includegraphics[width=1.0\linewidth, trim= 0 9.1cm 0 9.2cm, clip]{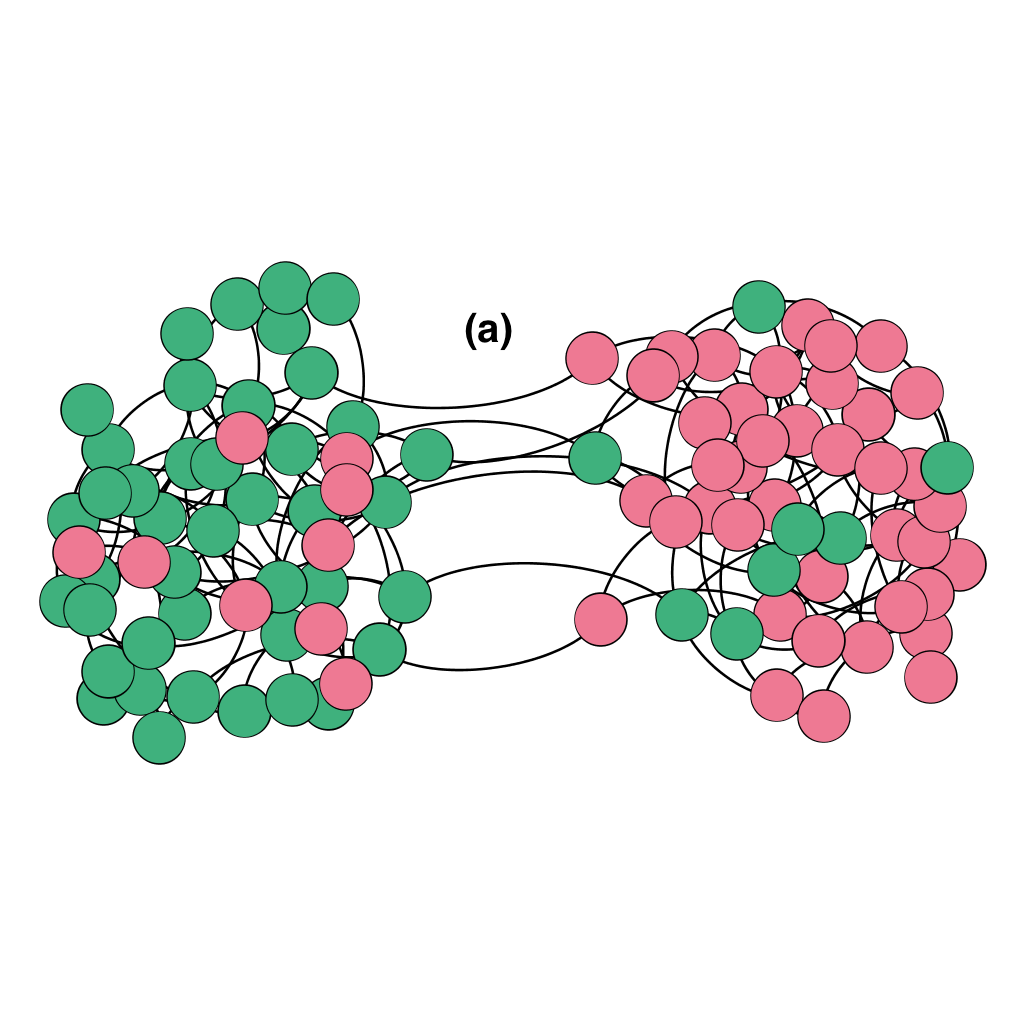} \label{fig:cdom1}
    }\vfill
    \subfloat{%
    \includegraphics[width=1.0\linewidth, trim= 0 9.1cm 0 9.2cm, clip]{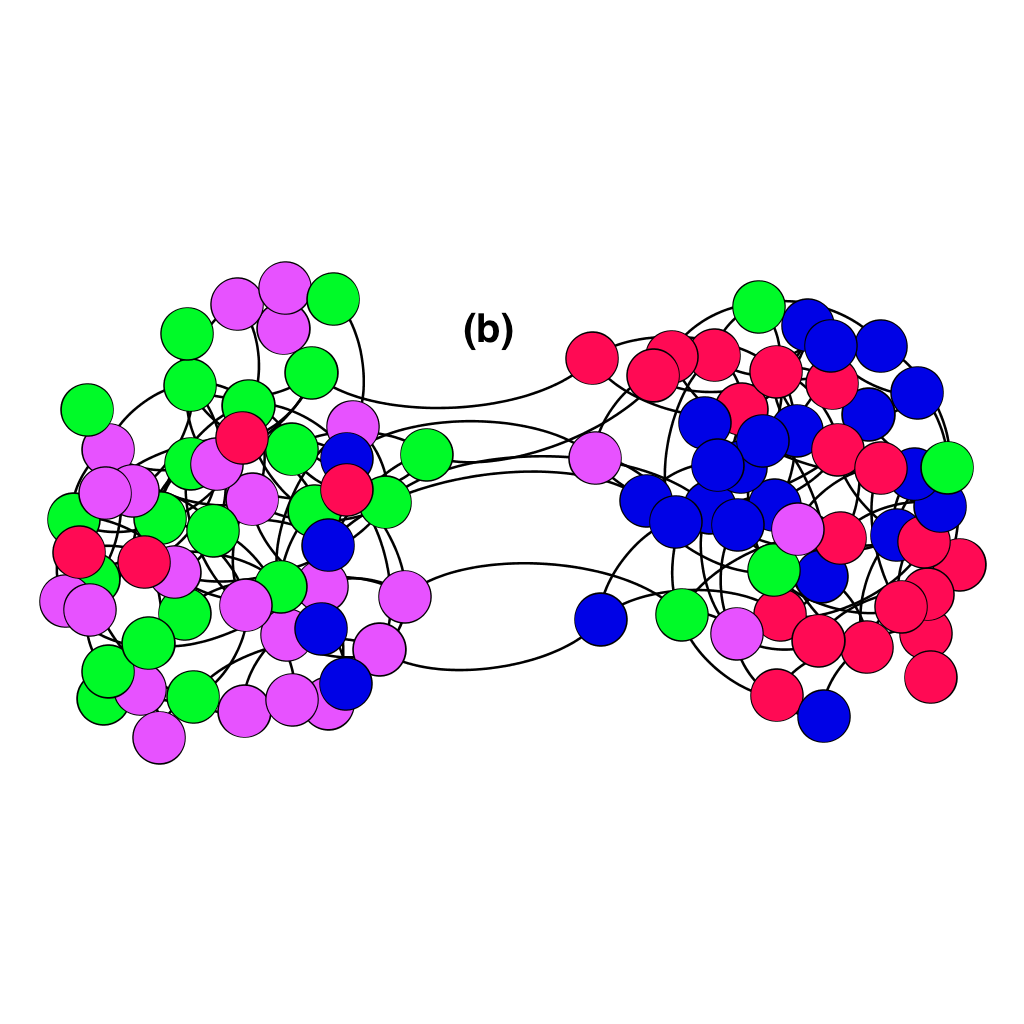} \label{fig:cdom2}
    }
	\vfill
	\caption{Illustration of the model (color online). In order for two nodes to be securely connected, they must have at least one path between them that avoids each color. In {\bf (a)} we show the case of two colors, $C=2$, with a single dominant color, $C_{d}=1$, in each module with $q=0.8$. In {\bf (b)} we demonstrate the case of $C_{d}=2$ (total number of colors, $C=4$) again with $q=0.8$. For this case each dominant color occupies only $q/C_{d}=40\%$ of its respective module. }
	\label{fig:c-colors}
\end{figure}

For simplicity, we demonstrate our model and results on a network with two communities having an internal average degree  $k_\text{I}$ and an external average degree $k_\text{E}$ \cite{SM-communities}. We begin by assuming (for simplicity but without loss of generalization) that there are two colors with a single dominant color ($C_d=1$) occupying a fraction $q$ nodes of each module and the remaining fraction $1-q$ being of the other color (see Fig \ref{fig:cdom1}) \cite{SM-hetero-q}. 
 This same framework can be used to describe networks where the links are correlated by color (See SM). To identify the giant secure component (GSC), we find the standard giant component under the removal of nodes of a single color, and then add back nodes of the removed color which have a direct link to the largest component (reflecting the assumption that the endpoints of every path are secure) \cite{krause2016hidden}. This is done for each color and then the intersection of all these components is the GSC. 

\begin{figure}
	\centering
	\includegraphics[width=1.0\linewidth]{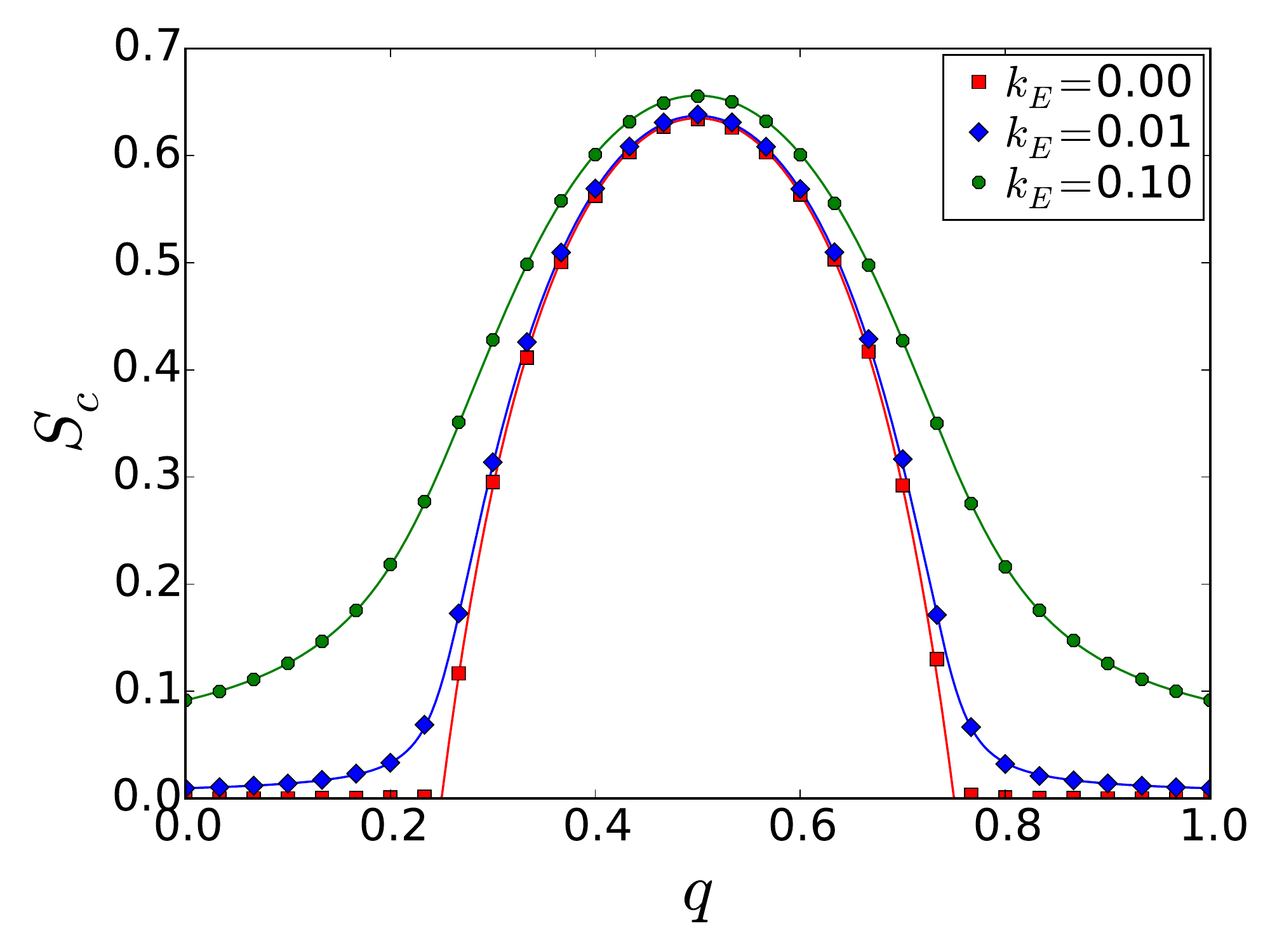}
	\caption{Normalized size of the GSC, $S_{c}$, as a function of dominance, $q$, for a network with 2 modules having \er structure, a single dominant color in each module, fixed $k_I=4$, and increasing levels of $k_{E}$. The lines, representing theory according to Eqs. (\ref{eq:s-c2}) and (\ref{eq:u-vals2}), show excellent agreement with simulations (symbols) on systems of size $N=10^6$ nodes. For the case $k_{E}=0$, we observe a phase transition as the level of dominance reaches the critical point $q_c=0.75$, while for non-vanishing $k_{E}$ no phase transition occurs. Due to the model symmetry for 2 modules, we also observe a transition at $1-q_c=0.25$.} 
	\label{fig:theory}
\end{figure}

To solve our model analytically, we adopt the generating function framework defining $g_0(z)=\sum_k p_k z^k$ as the generating function of the variable $k$ with $p_k$ being the probability of a node having $k$ links \cite{newman-book2009,callaway-prl2000}. 
For our model we have generating functions for the internal and external connections defined by $g_{0_{k_I}}(z)$ and $g_{0_{k_E}}(z)$ respectively. 
For the case of 2 colors, we must find: $u_{1,0}$, the likelihood that a link fails to avoid the color dominant in its module; $u_{0,1}$, the likelihood that the link fails to avoid the color dominant in the other module; and $u_{1,1}$, the likelihood that the link does not avoid either of the two colors. We then assume that the sender and receiver nodes are secure, by taking $g_{0_{k_I}}(u_{i,j})g_{0_{k_E}}(u_{j,i})$, which adds back nodes with a direct link to the giant component in both the internal and external modules.  
Naively one might think that to find the size of the GSC, $S_c$, one could merely take $1-g_{0_{k_I}}(u_{1,0})g_{0_{k_E}}(u_{0,1})-g_{0_{k_I}}(u_{0,1})g_{0_{k_E}}(u_{1,0})$ i.e., take the conjugate of the probability that a randomly chosen node fails to avoid both colors. However, this neglects the fact that some nodes fail to avoid either color. To deal with this overcounting we must add back $g_{0_{k_I+k_E}}(u_{1,1})$ in accordance with the inclusion-exclusion principle \cite{grinstead2012introduction}. Thus, we obtain
\begin{eqnarray}
 S_c&=&1-g_{0_{k_I}}(u_{1,0})g_{0_{k_E}}(u_{0,1}) \label{eq:s-c2}\\ \nonumber &-&g_{0_{k_I}}(u_{0,1})g_{0_{k_E}}(u_{1,0})+g_{0_{k_E+k_I}}(u_{1,1}).
\end{eqnarray}
To solve Eq.~\eqref{eq:s-c2} we need to calculate the probabilities $u_{i,j}$ which, for \er topologies of internal and external connections, are obtained by solving self-consistently the system
\begin{eqnarray}
u_{1,0}&=&q+(1-q)e^{-k_I(1-u_{1,0})-k_E(1-u_{0,1})} \nonumber\\
u_{0,1}&=&(1-q)+qe^{-k_I(1-u_{0,1})-k_E(1-u_{1,0})} \label{eq:u-vals2} \\
u_{1,1}&=&qe^{-k_I(1-u_{0,1})-k_E(1-u_{1,0})}+(1-q)e^{-k_I(1-u_{1,0})-k_E(1-u_{0,1})}. \nonumber
\end{eqnarray}
Results comparing the above theory to simulations are shown in Fig.~\ref{fig:theory}.

We find from Fig.~\ref{fig:theory} that only in the case where $k_{E}=0$ does the system undergo a phase transition  at the critical point $q_c=1-1/k_{I}$ \cite{krause2016color}, while for any $k_{E}>0$ there is always  some fraction of nodes in the secure component. 
This is because even if one of the two modules disintegrates when the dominant color is removed from it, there always exists a finite fraction of its nodes which can communicate securely through external links to the other module. Thus $k_E>0$ removes the transition by making the disconnected phase unreachable, just as an external magnetic field of magnitude $H$ does with respect to the disordered phase in the Ising model \cite{huang1963statistical}. In what follows we further support, both analytically and by extensive simulations, this intriguing analogy between spin models and secure message-passing on modular networks.\\
%

\begin{figure}
	\centering
	\hfill
		\subfloat{
		\includegraphics[width=0.48\linewidth]{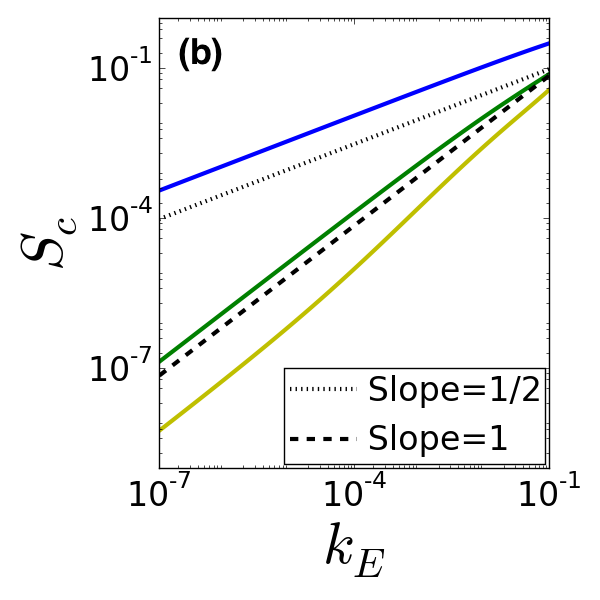} \label{fig:delta}
    }\hfill
    \subfloat{%
    \includegraphics[width=0.48\linewidth]{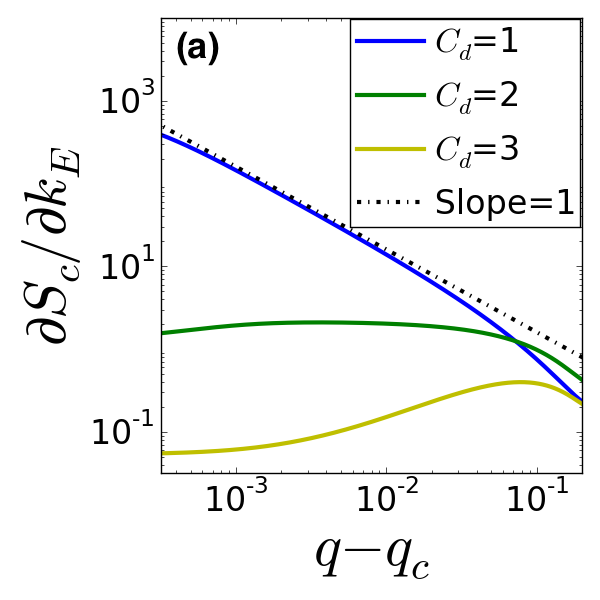} \label{fig:gamma}
    } \hfill
	\caption{ Critical scaling and higher-order transitions (color online).  {\bf (a)} Scaling of $S_c$ as a function of $k_E$ at the critical level of dominance $q_c=0.75$ with $k_{I}=C_{d}/(C_{d}-0.75)$. For $C_{d}=1$ we obtain $\delta=2$, whereas for $C_{d}>1$ we find $\delta=1$. {\bf (b)} Shown is the CAP analogue of the magnetic susceptibility near criticality as $k_{E} \to 0$. We take the difference between the curves for $S_{c}$ with $k_{E}=0$ and $k_{E}=10^{-6}$. For $C_{d}=1$ we find $\gamma=1$, whereas $\gamma=0$ for $C_{d}>1$. The latter result suggests that the system undergoes higher-order phase transitions \cite{janke2006properties} for more than two dominant colors.} 
	\label{fig:delta-gamma-exponents}
\end{figure}

We now investigate the scaling relations of our model with $S_{c}$, $q$, and $k_{E}$ as the CAP analogues of total magnetization, temperature, and the external field respectively. We note that for the case $C_d=1$, namely the case of a single dominant color in each module, the exponent $\beta$, defined by $S_c(q_c)\sim(q-q_c)^\beta$, is given by $\beta=1$ \cite{krause2016hidden}.  We will now measure $\delta$, which defines the scaling of the order parameter with the external field. In our analogy this is given by
\begin{equation}
S_{c}\sim k_{E}^{1/\delta}.
\label{eq:delta-defined}
\end{equation} 
For $C_d=1$ we obtain $\delta=2$ (Fig. \ref{fig:delta}). Thus, in this case, increasing external connectivity is more beneficial near the critical point since $1=\beta>1/\delta=\frac{1}{2}$. 

We next consider the analogue of the magnetic susceptibility, which satisfies the scaling relation
\begin{equation}
\left(\frac{\partial S_{c}}{\partial k_{E}}\right)_{k_{E}\to 0}\sim|q-q_c|^{-\gamma}.
\label{eq:gamma-defined}
\end{equation}
Using  Eqs.~\eqref{eq:s-c2} and \eqref{eq:u-vals2}, we find (Fig. \ref{fig:gamma}) $\gamma=1$ for $C_d=1$. We note that the exponents obtained ($\delta=2$, $\gamma=1$, and $\beta=1$) are consistent with Widom's identitiy $\delta-1=\gamma/\beta$ \cite{bunde1991fractals,*stanley1971phase}. 

The numerical results above can also be found analytically by expanding for $k_I$ near its critical value, $k_I=\frac{1}{1-q_c}$. By defining $x_{1,0}\equiv1-u_{1,0}$ and $x_{0,1}\equiv1-u_{0,1}$ and expanding Eq.~\eqref{eq:u-vals2} to leading orders in $x_{1,0}$ and $k_E$, we obtain
\begin{equation}
x_{1,0}=q_c-q+\sqrt{(q_c-q)^2+\frac{2k_Ex_{0,1}}{k_I^2}}.
\label{eq:scaling-theory}
\end{equation}
It follows that $\delta=2$, as $x_{1,0}$ scales with the square root of $k_E$, and $\gamma=1$ as can be found by taking the derivative of Eq.~\eqref{eq:scaling-theory} with respect to $k_E$. 

Having discussed the case of a single dominant color, we now study the case of multiple colors ($C_d>1$) sharing dominance in a single community as depicted in Fig. \ref{fig:cdom2}. Each of these dominant colors will occupy a fraction $q/C_d$ of the module. Following logic similar to that used for $C_d=1$, the GSC in this case can be found by
\begin{equation}
S_{c}=\sum_{i=0}^{C_{d}}\sum_{j=0}^{C_d}(-1)^{(i+j)}\binom{C_{d}}{i}\binom{C_{d}}{j}e^{-k_\text{I}(1-u_{i,j})-k_\text{E}(1-u_{j,i})}
\label{eq:S-general} 
\end{equation}
where the probabilities $u_{i,j}$ satisfy the system of self-consistent equations 
\begin{eqnarray}
u_{i,j}&=&i\frac{q}{C_{d}}e^{-k_\text{I}(1-u_{i-1,j})-k_\text{E}(1-u_{j,i-1})}\nonumber\\&&+j\frac{1-q}{C_{d}}e^{-k_\text{I}(1-u_{i,j-1})-k_\text{E}(1-u_{j-1,i})}\nonumber\\&&+\left(1-i\frac{q}{C_d}-j\frac{1-q}{C_{d}}\right)e^{-k_\text{I}(1-u_{i,j})-k_\text{E}(1-u_{j,i})} 
\end{eqnarray}
with $i\leq C_{d}$, $j\leq C_{d}$, and  $u_{0,0}=u_{0,-1}=u_{-1,0}\equiv1$. For $k_E=0$ we recover the equations of Krause et al. \cite{krause2016hidden,krause2016color}.

In contrast with the results for $C_d=1$, we find that for every $C_d\geq2$ the critical scaling exponents are given by $\gamma=0$ and $\delta=1$ (Fig. \ref{fig:delta-gamma-exponents}) which define a novel universality class.
These results, together with the exponent $\beta$, in this case given by $\beta=C_d$ \cite{krause2016color}, suggest that for more than one dominant color the system undergoes {\em higher}-{\em order} phase transitions. To verify this claim, we evaluate the higher-order derivatives of $S_c$ with respect to $k_E$, the first of which is given by
\begin{equation}
\left(\frac{\partial^{2}S_c}{\partial k_E^{2}}\right)_{k_E\to0}\sim \big(q-q_c\big)^{-G},
\label{eq:G-defined}
\end{equation}
\noindent
where $G$ satisfies the generalized scaling relation $G=\beta\big(C_d\delta-1\big)$ \cite{janke2006properties}. In particular, for $C_d=2$ we expect an exponent $G=2$, which we confirm with numerical results (see SM). For $C_d\geq3$, Eq.~\eqref{eq:G-defined} breaks down and we obtain $G=1$. To the best of our knowledge the present study represents the first time that this novel universality class with higher-order transitions is observed in percolation type systems with the higher-order scaling exponents defined and measured. 

\begin{figure}
	\centering
	\includegraphics[width=1.0\linewidth]{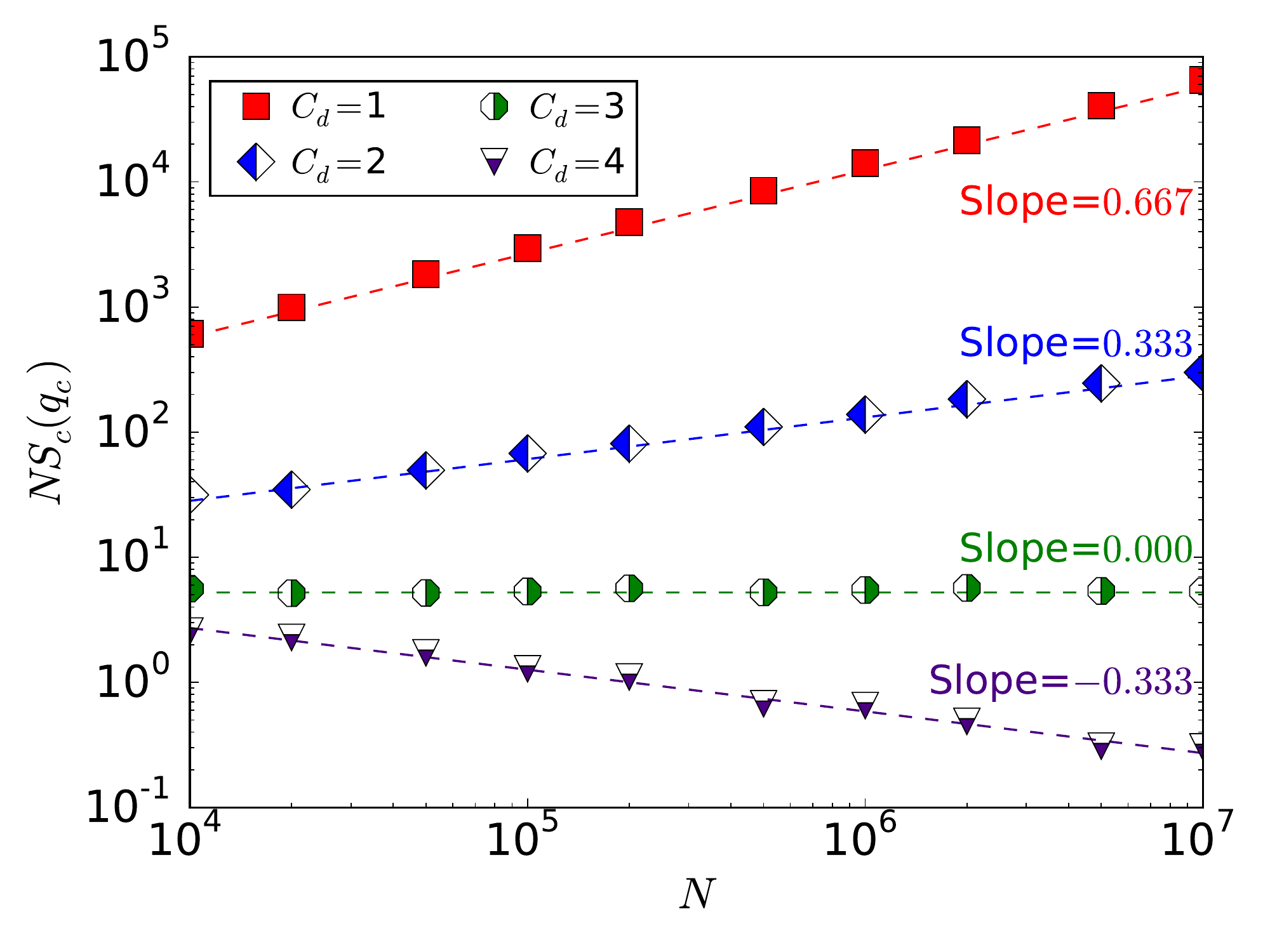}
	\caption{Size of the secure component at criticality (color online). The points represent averages over at least $400$ simulations, while the dashed lines represent slopes of $1-C_{d}/3$  as predicted in Eq.~\eqref{eq:scaling-at-pc}. For all $C_{d}$ we observe excellent agreement between these predictions and the simulations.  }
	\label{fig:nu-dc}
\end{figure}

Finally, we can indirectly evaluate $\nu d_c$, where $\nu$ is the scaling exponent of the correlation length at criticality and $d_c$ is the upper critical dimension of CAP \cite{coniglio1982cluster}. We do this by analyzing how the size of the GSC, $NS_{c}(q_c)$, scales at criticality with the number of nodes $N$ in the absence of external connections (i.e., $k_{E}=0$). This gives us $\nu d_c$ as follows: We know \cite{bunde1991fractals,*stanley1971phase} that $S_c\sim (q-q_c)^\beta$, the correlation length $\xi \sim |q-q_c|^{-\nu}$ near criticality, and $N\sim\xi^{d_c}$ at criticality. Combining all  these gives $S_c\sim N^{-\beta/\nu d_c}$ or equivalently $NS_{c}(q_c)\sim N^{1-\beta/\nu d_c}$. Recalling that $\beta=C_{d}$ \cite{krause2016hidden}, by measuring $NS_{c}(q_c)$ for varying $N$, we can find $\nu d_c$. In Fig. \ref{fig:nu-dc} we carry out this simulation for different $C_d$ and obtain $\nu d_c=3$, most likely with $\nu=\frac{1}{2}$ and $d_c=6$ as for classical percolation on \er networks. 

This can be understood as follows: The scaling of $S_c(q_c)\sim N S_1(q_c)  S_2(q_c)... S_{C_d}(q_c)=\frac{N}{N^{C_d}} NS_1(q_c) \times NS_2(q_c)...  \times NS_{C_d}(q_c)$, where $S_1(q_c),...,S_{C_d}(q_c)$ represent the scaling of the size of the component avoiding color $1...C_d$. Each $S_1(q_c),...,S_{C_d}(q_c)$ scales like an \er network \cite{krause2016hidden} with $NS_1(q_c),...,NS_{C_d}(q_c)\sim N^{2/3}$. If we rearrange and substitute this into our expression above we obtain $S_c(q_c)\sim \frac{N}{N^{C_d}} N^{2/3}\times N^{2/3}... \times N^{2/3}$ and finally
\begin{equation}
 S_c(q_c)\sim N^{1-C_d} N^{\frac{2C_d}{3}}=N^{1-C_d/3}.
\label{eq:scaling-at-pc}
\end{equation}
 This can then be set equal to $N^{1-\beta/\nu d_c}$ to obtain the numerical result $\nu d_c=3$. 

This constant value of $\nu d_c$ combined with the increasing value of $\beta$ as the number of colors increases, leads to the rather surprising behavior of Fig.~\ref{fig:nu-dc} where the size of the largest cluster, $NS_{c}(q_c)$, \emph{decreases} with the system size $N$, when $C_{d}>3$. We explain this effect by noting that for a node to be in the secure component it must be in the intersection of the components avoiding each color. The likelihood of avoiding any single color scales with $N^{-1/3}$, such that when two colors must be avoided the scaling is $N^{-1/3} \times N^{-1/3}$, and so on for additional colors. Once more than three colors must be avoided, the decreasing likelihood of being in all of the colors overpowers the linear growth of the system size leading to the observed decrease. Further, this suggest that at criticality the system has vanishing fractal dimension for $C_d=3$, and increasingly negative fractal dimension for $C_d>3$ \cite{mandelbrot-physa1990}. We stress that the surprising values of vanishing and negative fractal dimension is unprecedented in the context of percolation on networks. For instance, in classical percolation on scale-free networks, $\beta$ increases as the degree-distribution becomes broader \cite{cohen2003structural,burda2001statistical}, but this increase is counteracted by the simultaneous increase of the upper critical dimension, thus the fractal dimension remains positive. 


Finally, our results also suggest the breakdown of the scaling relation $\nu d_c=2\beta+\gamma$ \cite{bunde1991fractals,*stanley1971phase} for $C_{d}>1$ since $\nu d_c=3$ for all $C_{d}$ but $2\beta+\gamma=2C_{d}$ (for $C_d>1$) which increases with $C_{d}$. This scaling relation originated from the distribution of finite clusters at criticality scaling with an exponent $\tau<3$ \cite{bunde1991fractals,*stanley1971phase}. Its failure here implies that for color-avoiding percolation with $C_{d}>1$, $\tau\geq3$. This can be understood based on previous results on the bicomponent, which is less restrictive than color-avoiding percolation, where it was shown that there are no large finite size bicomponents, rather only a giant bicomponent can exist \cite{newman2008bicomponents}.

In summary, our results maps the study of secure message passing between nodes in modular networks to the statistical physics of Ising models with a magnetic field.  Previous attempts to introduce the idea of a field into percolation relied on a ghost site \cite{reynolds1977ghost}, to which every node connects with some probability $H$ and thus allowing it to remain functional even if it is separated from the `rest' of the largest cluster.  Here we obtain the field exponents, $\delta$ and $\gamma$, naturally as a result of the realistic effects of modules rather than from the artificial introduction of a ghost site. Further, we find novel universality classes, the breakdown of a known scaling relation and higher-order phase transitions. This work highlights the potential for incorporating the idea of an external field into complex systems and shows how this idea can be used to shed light on the fundamental physics of the system.

We acknowledge the Israel-Italian collaborative project NECST, Israel Science Foundation, ONR, Japan Science Foundation, BSF-NSF, and DTRA (Grant no. HDTRA-1-10-1- 0014) for financial support. GC acknowledges support from EU projects SoBigData nr. 654024 and CoeGSS nr. 676547. S.B. acknowledges the B. W. Gamson Computational Science Center at Yeshiva College. V.Z. acknowledges support by the H2020 CSA Twinning Project No. 692194, RBI-T-WINNING, and Croatian centers of excellence QuantixLie and Center of Research Excellence for Data Science and Cooperative Systems.
\bibliographystyle{naturemag}
\bibliography{paper}

\end{document}